\documentstyle[aps,prl,preprint,tighten]{revtex}
\input{psfig}
\begin{document}
\draft
\preprint{LU TP 96-22}
\title{Relation between a Screened Polyelectrolyte and a Field Theory}
\author{Bo S\"{o}derberg}
\address{Department of Theoretical Physics, University of Lund,
	S\"{o}lvegatan 14A, S-223 62 Lund, Sweden}
\maketitle
\begin{abstract}
There exists a generic relationship between the thermodynamics of a
continuous polymer with a generic self-interaction and the two-point
function of an interacting field-theory. In addition, the $(2N)$-point
function of the resulting field theory is similarly related to a
system of $N$ interacting polymers.

In the present paper, this relation is explored for the special case
of a polyelectrolyte, characterized by a screened Coulomb pair
potential. The corresponding field theory can be recast in a
particularly simple form, corresponding to a quantum-mechanical
particle, self-interacting via the emission and absorption of a
massive scalar field.

This is particularly useful in a perturbative treatment: perturbative
expansions for the polyelectrolyte can be simply derived from the loop
expansion for the related field-theoretical two-point function, for
which established computational methods exist.
\end{abstract}

\pacs{PACS numbers: 05.70.-a, 36.20.Ey, 11.10.C}

\narrowtext

\paragraph*{Introduction}

Relations between Euclidean field theories and random walks or
polymers are well-known
\cite{symanzik,degennes,degennes2,parisi,itzykson,CJ}. For a
non-interacting continuous polymer, equivalent to a random walk, the
relation to the propagator of a free field theory is trivial. For the
case of a continuous polymer with a delta-function pair interaction,
corresponding to a self-avoiding random walk, de Gennes has shown the
relation to a $\phi^4$ field theory (Landau-Ginsburg model)
\cite{degennes,degennes2,parisi,itzykson}. A similar relation has been
exploited for an unscreened polyelectrolyte \cite{pfeuty}. The focus
of the present paper is to exploit the generic result for the special
case of a screened polyelectrolyte, leading to a specific, very simple
field-theory.

\paragraph*{The generic interacting polymer}

A quite generic self-interacting discrete polymer with $N$ monomers in
$D$-dimensional space is described by the Hamiltonian
\begin{eqnarray} \nonumber & &
	H \left ( {\bf x}_1, \ldots, {\bf x}_N \right ) =
	\frac{1}{2} \sum_{i=1}^{N-1} \left ( {\bf x}_{i+1} - {\bf x}_i
	\right )^2 +
	\sum_{i=1}^N \tilde{V}_1 \left ( {\bf x}_i \right ) +
\\ \nonumber & &
	\sum_{i=1}^N \sum_{j=i+1}^N \! \tilde{V}_2 \left ( {\bf x}_i,{\bf x}_j
	\right )
	+ \sum_{i=1}^N \sum_{j=i+1}^N \sum_{k=j+1}^N
	\! \tilde{V}_3 \left ( {\bf x}_i,{\bf x}_j,{\bf x}_k \right ) + \ldots,
\end{eqnarray}
where ${\bf x}_i$ denotes the position in ${\cal R}^D$ of the $i$th
monomer. The first term represents a harmonic bonding force between
nearest-neighbour monomers, while $\tilde{V}_1({\bf x})$ is a generic
one-particle potential, $\tilde{V}_2({\bf x},{\bf y})$ a ditto pair
potential, etc. The interaction is the most general one that depends
on positions only and is independent of monomer indices. Thermodynamic
averages at temperature $T$ are defined based on the Boltzmann
distribution, $P\left ({\bf x}_1,\ldots,{\bf x}_N\right ) \propto
\exp\left (-H/T\right )$.

For the Boltzmann exponent $S \equiv H/T$, we obtain in the continuum
approximation, defined by replacing the discrete monomer index $i$ by
a continuous chain parameter $\tau = Ti$, such that ${\bf x}_i \to
{\bf x}(Ti) \equiv {\bf x}(\tau)$, the following expression:
\begin{eqnarray}
\nonumber &&
	S[{\bf x}(\tau)] =
	\frac{1}{2} \int_0^{\mu} d\tau \; \dot{{\bf x}}^2
	+ \int_0^{\mu} d\tau \; V_1\left ({\bf x}(\tau)\right ) +
\\ \nonumber &&
	\; \frac{1}{2} \int_0^{\mu} d\tau \int_0^{\mu} d\tau' \;
	V_2\left ({\bf x}(\tau),{\bf x}(\tau')\right ) +
\\ \nonumber &&
	\; \frac{1}{6}
	\int_0^{\mu} d\tau \int_0^{\mu} d\tau' \int_0^{\mu} d\tau''
	\; V_3\left ({\bf x}(\tau),{\bf x}(\tau'),{\bf x}(\tau'')\right )
	+ \ldots ,
\end{eqnarray}
where square brackets $[\,]$ are used to indicate that $S$ is a
functional. The length of the polymer is represented by $\mu=NT$,
while $V_k \equiv \tilde{V}_k/T^{k+1}$, for $k = 1,2,\ldots$. The dot
in $\dot{{\bf x}}$ stands for the derivative with respect to
$\tau$. The continuum approximation is valid when the
nearest-neighbour distances are small in comparison to the range of
the interactions.  The independence on monomer index is turned into a
translation invariance with respect to the chain position $\tau$:
$V_n$ depends on $(\tau,\tau',\ldots)$ only via $({\bf x}(\tau),{\bf
x}(\tau'),\ldots)$.

The continuum formulation will be the starting point for the analysis
carried out in the rest of this letter, and $S[{\bf x}(\tau)]$ will be
referred to as the {\em action} of the system.  To begin with, we will
assume fixed B.C., with the endpoints of the chain locked at ${\bf
x}(0)={\bf a}$, ${\bf x}(\mu)={\bf b}$. The main object of interest is
the {\em partition function}, which is given by the functional
integral
\[
	Z(\mu,{\bf a},{\bf b}) =
	\int_{{\bf x}(0)={\bf a}}^{{\bf x}(\mu)={\bf b}}
	d[{\bf x}(\tau)] \exp\left ( -S[{\bf x}(\tau)]\right ) .
\]
The proper partition function for free boundary conditions is
obtained simply by integrating $Z(\mu,{\bf a},{\bf b})$ over the
endpoints ${\bf a},{\bf b}$.

\paragraph*{The relation to a field theory}

The generic relation to a field-theory is has been described in the
literature (see e.g. \cite{CJ} and references therein). For the sake
of completeness, we will here provide a derivation.

The translation-invariance of the interaction ensures that the
interaction part $S_{\mbox{\footnotesize{int}}}$ of the action be a
functional only of the {\em particle density}, $ \rho({\bf y}) =
\int_0^{\mu} d\tau \delta\left ({\bf y} - {\bf x}(\tau)\right ) $, in
terms of which it is given by
\begin{eqnarray}
\nonumber &&
	\int d{\bf x} \rho({\bf x}) V_1({\bf x})
	+ \frac{1}{2} \int d{\bf x} \rho({\bf x})
	\int d{\bf y} \rho({\bf y}) V_2({\bf x},{\bf y}) +
\\ \nonumber &&
	\; \frac{1}{6} \int d{\bf x} \rho({\bf x}) \int d{\bf y} \rho({\bf y})
	\int d{\bf z} \rho({\bf z}) V_3({\bf x},{\bf y},{\bf z})
	+ \ldots
\end{eqnarray}
as a functional Taylor expansion. Next, we assume that the
corresponding Boltzmann factor, $\exp(-S_{\mbox{\footnotesize{int}}})$, can be
functionally Laplace transformed with respect to the (non-negative)
particle density,
\[
	F[\xi({\bf x})] = \int d[\rho({\bf x})] \exp\left (-\int d{\bf
	x} \xi({\bf x}) \rho({\bf x}) -
	S_{\mbox{\footnotesize{int}}}[\rho] \right ) .
\]
This is well-defined for any combination of arbitrary one-particle
potentials and non-negative many-particle potentials. The functional
Laplace transform can be inverted to yield, formally,
\[
	\exp\left (- S_{\mbox{\footnotesize{int}}}[\rho]\right )
	\propto \int d[\xi({\bf x})] F[\xi({\bf x})] \exp\left (\int
	d{\bf x} \xi({\bf x}) \rho({\bf x}) \right ) ,
\]
where each $\xi({\bf x})$ is to be integrated along a complex contour
to the right of all singularities. We then obtain for the partition
function
\begin{eqnarray}
\nonumber &&
	Z(\mu,{\bf a},{\bf b}) \propto
	\int d[\xi({\bf x})] F[\xi({\bf x})] \times
\\ \nonumber &&
	\int_{{\bf x}(0)={\bf a}}^{{\bf x}(\mu)={\bf b}}
	d[{\bf x}(\tau)] \exp\left (
	-\frac{1}{2} \int_0^{\mu} d\tau \; \dot{{\bf x}}^2
	+ \int_0^{\mu} d\tau \xi({\bf x}(\tau))
	\right ) ,
\end{eqnarray}
and $-\xi({\bf x})$ appears as an effective one-particle potential for
the polymer.  Now the ${\bf x}$ integral is recognized as the
Euclidean quantum-mechanical path integral for the amplitude $< {\bf
a} | \exp ( -\mu \hat{H} ) | {\bf b} >$ to go from ${\bf a}$ to ${\bf
b}$ in imaginary time $\mu$, with the quantum-mechanical Hamiltonian
\[
	\hat{H} = -\frac{1}{2} \nabla^2 - \xi({\bf x}) + \mbox{\it const.} .
\]
We thus have
\[
	Z(\mu,{\bf a},{\bf b}) \propto
	\int d[\xi({\bf x})] F[\xi({\bf x})]
	\left < {\bf a} \left |
	\exp \left ( -\mu \hat{H} \right )
	\right | {\bf b} \right > .
\]
Now, we can make a Laplace transform in $\mu$,
\[
	W(s,{\bf a},{\bf b}) = \int_0^{\infty} d\mu \exp(-s\mu)
	Z(\mu,{\bf a},{\bf b}) ,
\]
yielding
\[
	W(s,{\bf a},{\bf b}) \propto
	\int d[\xi({\bf x})] F[\xi({\bf x})]
	\left < {\bf a} \left |
	\left ( \hat{H} + s \right )^{-1}
	\right | {\bf b} \right > .
\]
The matrix-element can be written as a functional integral over
a field $\vec{\phi}({\bf x})$ having $n=0$ components, 
\begin{eqnarray}
\nonumber &&
	\left < {\bf a} \left |
	\left ( \hat{H} + s \right )^{-1}
	\right | {\bf b} \right >
	=
\\ \nonumber &&
	2
	\int d[\vec{\phi}({\bf x})]
	\phi_1({\bf a}) \phi_1({\bf b}) \times
	\; \exp \left ( -
	\int d{\bf x}
	\vec{\phi} \left ( \hat{H} + s \right ) \vec{\phi}
	\right ) ,
\end{eqnarray}
and we obtain
\begin{eqnarray}
\nonumber &&
	W(s,{\bf a},{\bf b}) \propto
	\int d[\xi({\bf x})] F[\xi({\bf x})]
	\int d[\vec{\phi}({\bf x})]
	\phi_1({\bf a}) \phi_1({\bf b}) \times
\\ \nonumber &&
	\; \exp \left (-
	\int d{\bf x} \left \{
	\frac{1}{2} \left (\nabla \vec{\phi}\right )^2
	+ s \vec{\phi}^2
	- \xi({\bf x}) \vec{\phi}^2
	\right \}
	\right ) .
\end{eqnarray}
Now we can do the $\xi$ integral to recover $\exp\left
(-S_{\mbox{\footnotesize{int}}}[\rho({\bf x})]\right )$, but with a
modified particle density, $ \rho({\bf x}) \to \vec{\phi}({\bf x})^2 $.
Thus, we finally arrive at the following result:
\[
	W(s,{\bf a},{\bf b}) \propto
	\int d[\vec{\phi}({\bf x})]
	\phi_1({\bf a}) \phi_1({\bf b})
	\exp \left ( -\hat{S}[\phi({\bf x})] \right ) .
\]
The RHS defines the two-point function $G(s,{\bf a},{\bf b})$ of an
interacting field theory, with an action $\hat{S}$ given by
\begin{eqnarray}
\nonumber
	& & \hat{S}[\phi({\bf x})] =
	\int d{\bf x} \left \{
	\frac{1}{2} \left (\nabla \vec{\phi}\right )^2
	+ s \vec{\phi}^2({\bf x})
	+ V_1({\bf x}) \vec{\phi}^2({\bf x})
	\right \} +
\\ \nonumber &&
	\; \frac{1}{2} \int d{\bf x} \int d{\bf y} \; V_2({\bf x},{\bf y})
	\vec{\phi}^2({\bf x})
	\vec{\phi}^2({\bf y}) +
\\ \nonumber &&
	\; \frac{1}{6} \int d{\bf x} \int d{\bf y} \int d{\bf z}
	\; V_3({\bf x},{\bf y},{\bf z})
	\vec{\phi}^2({\bf x}) \vec{\phi}^2({\bf y}) \vec{\phi}^2({\bf z})
	+ \ldots
\end{eqnarray}

This shows that the partition function for a generic interacting
polymer, when Laplace transformed with respect to the chain-length
$\mu$, yields the propagator of a zero-component field theory, with a
bare mass determined by the corresponding conjugate variable $s$, and
where the interaction appears in terms of a particle density given by
the square of the field. Upon collecting the neglected constants of
proportionality accumulated in the tranformations, the exact relation
is given by
\[
	2 G(s,{\bf a},{\bf b}) = \int_0^{\infty} d\mu \exp(-s \mu)
	\frac{Z(\mu,{\bf a},{\bf b})}{\int d{\bf a} Z_0(\mu,{\bf a},{\bf 0})} 
\]
where $Z_0$ denotes the partition function of a non-interacting
polymer; the correctness can be verified by considering the case of no
interactions, since the constants of proportionality involved do not
depend on the interaction.

\paragraph*{Several polymers}

Having established the relation between the partition function of an
interacting polymer and the two-point function of a related field
theory, one might wonder what is the relevance of the higher order
moments in $\vec{\phi}$. It turns out (the proof is analogous to the
one-polymer case) that the four-point function
\[
	\left < \phi_1({\bf a}) \phi_1({\bf b}) \phi_2({\bf c})
	\phi_2({\bf d}) \right >
\]
is proportional to the Laplace transform of the partition function for
{\em two polymers}, one with its endpoints in ${\bf a}$ and ${\bf b}$,
the other in ${\bf c}$ and ${\bf d}$, interacting both internally and
with each other, and allowed to exchange monomers with each other: The
parameter $\mu$, conjugate to $s$, merely fixes their {\em total
length}. In addition, the moment
\[
	\left < \phi_1({\bf a}) \phi_1({\bf b}) \phi_1({\bf c})
	\phi_1({\bf d}) \right >
\]
corresponds to the sum of the three possible endpoint rearrangements
of the two polymers. Similarly, the six-point functions are related to
a system of three polymers, etc.

In contrast, a system of $N$ polymers of {\em fixed lengths} $\mu_i$
corresponds to, upon Laplace transforming each length individually, a
field theory with distinct squared masses $2s_i$ for $N$ subsets (of
zero components each) of the field components.

For dominantly repulsive interactions, the consideration of several
polymers will of course make sense only in a finite volume (or finite
polymer concentration).

\paragraph*{The screened polyelectrolyte}

In the physically interesting case of a screened polyelectrolyte in
three dimensions \cite{fake,scale}, having only a pair interaction
given by a screened Coulomb potential,
\[
	V_2({\bf x},{\bf y}) = g \exp(-\kappa r)/4\pi r, \; \mbox{with}
	\; r=|{\bf x}-{\bf y}| ,
\]
we have the resulting action
\begin{eqnarray} \nonumber &&
	\hat{S}[\phi({\bf x})] =
	\int d{\bf x} \left \{
	\frac{1}{2} \left (\nabla \vec{\phi}\right )^2
	+ s \vec{\phi}^2({\bf x})
	\right \} +
\\ \nonumber &&
	\; \frac{g}{2} \int d{\bf x} \int d{\bf y} \;
	\frac{\exp(-\kappa |{\bf x}-{\bf y}|)}{4\pi |{\bf x}-{\bf y}|}
	\vec{\phi}^2({\bf x}) \vec{\phi}^2({\bf y}) .
\end{eqnarray}
Noting that
\[
	\frac{\exp(-\kappa |{\bf x}-{\bf y}|)}{4\pi |{\bf x}-{\bf y}|} =
	\left ( -\nabla^2 + \kappa^2 \right )^{-1}({\bf x},{\bf y}) ,
\]
we can make a further simplification by introducing an additional
scalar field, the exchange of which mediates the interaction. This
gives
\begin{eqnarray} \nonumber
	W(s,{\bf a},{\bf b}) & \propto &
	\int d[\vec{\phi}({\bf x})]
	\int d[\psi({\bf x})]
	\phi_1({\bf a}) \phi_1({\bf b}) \times
\\ \nonumber &&
	\exp \left ( -\tilde{S}[\phi({\bf x}),\psi({\bf x})] \right ) ,
\end{eqnarray}
with a {\em local} combined action $\tilde{S}$ given by
\begin{eqnarray}
\nonumber &&
	\tilde{S}[\phi({\bf x}),\psi({\bf x})] =
\\ \nonumber &&
	\int d{\bf x} \left \{
	\frac{1}{2} \left (\nabla \vec{\phi}\right )^2
	+ s \vec{\phi}^2
	+ \frac{1}{2} \left (\nabla \psi\right )^2
	+ \frac{1}{2} \kappa^2 \psi^2
	+ i \sqrt {g} \vec{\phi}^2 \psi
	\right \} .
\end{eqnarray}
The variable $s$, conjugate to the chain length $\mu$, appears only in
the (bare) $\vec{\phi}$ mass, while the inverse range $\kappa$ of the
interaction potential appears as the $\psi$ mass; for the case of an
unscreened potential, the $\psi$-field will be massless.

Now we can modify the boundary conditions; noting that the polymer is
translationally and rotationally invariant, the partition function
$Z(\mu,{\bf a},{\bf b})$ can depend only on $R \equiv |{\bf a}-{\bf
b}|$. Integrating also over ${\bf R}$ would yield the relevant
partition function for the case of free boundary conditions. However,
it adds no complication to include a factor $\exp(i{\bf P} \cdot {\bf
R})$ in the integral, yielding
\begin{eqnarray}
\nonumber
	W(s,{\bf P}) &=&
	\int d{\bf R} \int_0^{\infty} d\mu \; \exp(-s\mu + i{\bf P}
	\cdot {\bf R}) Z(\mu,0,{\bf R})
\\ \nonumber
	&=&
	\int_0^{\infty} d\mu \exp(-s\mu)
	\tilde{Z}(\mu,{\bf P})
\\ \nonumber
	& \propto & \int d{\bf R} \exp(i{\bf P}\cdot{\bf R}) G(0,{\bf R})
	\equiv \tilde{G}({\bf P}) .
\end{eqnarray}
This yields the two-point function in momentum space, at momentum
${\bf P}$, for the $\vec{\phi}$ field. It contains all relevant information
on the distribution of the end-to-end distance ${\bf R}$ of the
polymer. Thus, eg., the average end-to-end distance squared,
$\left<R^2\right>$, can be obtained from
\[
	\frac{Z(\mu,{\bf P})}{Z(\mu,0)} = 1 - \frac{P^2\left<R^2\right>}{6}
	+ {\cal O}(P^4) .
\]

\paragraph*{Relation in terms of perturbation theory}

A perturbative (in the coupling $g$) expansion for the polyelectrolyte
is obtained by applying an inverse Laplace transform to the loop
expansion for the resulting local field theory, the latter amounting
to the well-known computation of Feynman graphs.  In three dimensions,
no ultraviolet divergences will appear, as can be seen from a simple
dimensional analysis.

The Feynman rules for computing $\tilde{G}({\bf P})$ are simple: On a
single $\phi$ line add lines of emitted and re-absorbed $\psi$ lines
joining the first line in vertices. A $\phi$ propagator carries a
factor $1/(q^2+2s)$ and a $\psi$ propagator a factor
$1/(k^2+\kappa^2)$. For every vertex add a factor $-2i\sqrt{g}$. The
external $\phi$ momentum is given by ${\bf P}$, and momentum is conserved
at every vertex. For every loop, add a momentum integral $\int
d^3p/(2\pi)^3$. It is easy to see from dimensional counting, that in
three dimensions the result will be ultra-violet finite for all
graphs.  Note that no $\phi$-loops are allowed (due to $n=0$); this
can be interpreted as the $\phi$-field corresponding to a
quantum-mechanical particle that is not second-quantized; thus, there
can be no creation or annihilation of the corresponding particles.

\begin{figure}
\centering \mbox{\psfig{figure=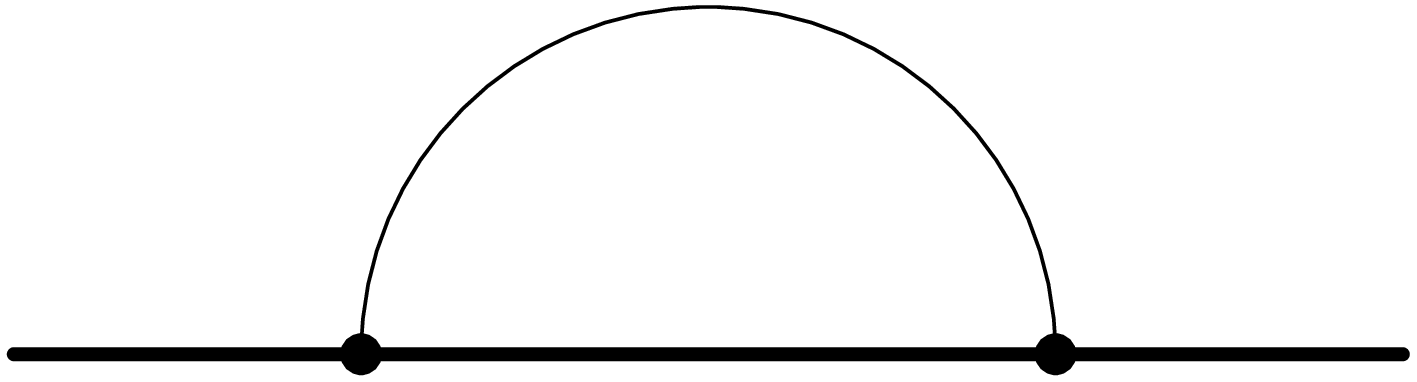,width=5cm}}
\caption{The only graph contributing to first order in the
perturbation expansion for $\tilde{G}({\bf P})$.}
\label{fig1}
\end{figure}

\begin{figure}
\centering \mbox{\psfig{figure=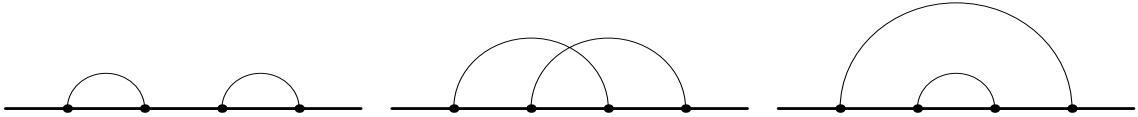,width=15cm}}
\caption{The three graphs contributing to second order in the
perturbation expansion for $\tilde{G}({\bf P})$.}
\label{fig2}
\end{figure}

Thus, to lowest order $\tilde{G}({\bf P})$ is given by the bare propagator
\[
	\tilde{G}^{(0)}({\bf P}) = \frac{1}{P^2+2s} .
\]
which is indeed half of the Laplace transform of $\exp(-\mu P^2/2)$,
reflecting the distribution of ${\bf R}$ for a free polymer being
proportional to $\exp(-{\bf R}^2/2\mu)$. To first order in $g$,
corresponding to one loop, there is only one graph contributing,
displayed in Fig.\ \ref{fig1}, yielding
\begin{eqnarray}
\nonumber
	\tilde{G}^{(1)}({\bf P})
	&=& \frac{-4g}{(P^2+2s)^2}
	\int \frac{d^3{\bf q}}{(2\pi)^3} \frac{1}{(q^2+\kappa^2)
	(({\bf P}-{\bf q})^2+2s)}
\\ \nonumber
	&=& \frac{-4g}{(P^2+2s)^2}
	\frac{1}{4\pi P} \arctan\left (\frac{P}{\sqrt{2s}+\kappa}\right ) .
\end{eqnarray}
To second order there are three distinct graphs, displayed in Fig.\
\ref{fig2}.

To compute the partition function $Z$, and related entities like
$\left< S_{\mbox{\footnotesize{int}}}\right>$, it is enough to
consider ${\bf P}=0$, while for $\left < R^2 \right >$, the leading
small $P^2$ corrections must be taken into account. Work on
perturbative calculations for the polyelectrolyte relying on the above
analysis is in progress, and will be presented elsewhere \cite{bs}.

In a similar way, any pair-potential that only depends on the relative
position, $V_2({\bf x},{\bf x}') = U({\bf x}-{\bf x}')$, can be
perturbatively treated; provided the $\psi$ propagator is replaced by
the Fourier transform of the potential, $\tilde{U}({\bf q}) = \int
d{\bf x} \exp(i{\bf q}\cdot{\bf x}) U({\bf x})$.

\newpage

\end{document}